\documentstyle[aps,prb,graphicx]{revtex}
\begin{document}
\draft
\twocolumn[\hsize\textwidth\columnwidth\hsize\csname 
@twocolumnfalse\endcsname
\title{Breakdown of the Landau--Fermi liquid in Two Dimensions due
  to Umklapp Scattering}
\author{C. Honerkamp$^{(1)}$, M. Salmhofer$^{(2)}$, N. Furukawa$^{(1,3)}$, and
T.M. Rice$^{(1)}$}
\address{$^{(1)}$ Theoretische Physik,
ETH-H\"onggerberg, CH-8093 Z\"urich, Switzerland,\\
$^{(2)}$ Mathematik, ETH Zentrum, CH-8092 Z\"urich, Switzerland, \\
$^{(3)}$ Department of Physics, Aoyama Gakuin University, Setagaya, Tokyo
157-8572, Japan\cite{presentadress}.}
\date{21 December 1999}
\maketitle 
\begin{abstract}
We study the renormalization group flow of the interactions in the two-dimensional $t$-$t'$ 
Hubbard model near half filling in a $N$-patch representation of
the whole Fermi surface. Starting from weak to intermediate couplings 
the flows are to strong coupling with different character depending on the 
choice of parameters. In a large parameter region elastic Umklapp
scatterings drive an instability which on parts of the Fermi surface exhibits the key signatures of an insulating spin liquid (ISL), as proposed by Furukawa et 
al., rather than a 
conventional symmetry-broken state. The ISL is characterized by 
both strong $d$-wave pairing and antiferromagnetic correlations, 
however it is insulating due to the vanishing local charge compressibility and a 
spin liquid because of the spin gap arising from the pairing 
correlations.  We find that the ISL is a consequence of 
a Fermi surface close to the saddle points at the 
Brillouin zone boundaries which provides an intrinsic and mutually
reinforcing coupling between 
pairing and Umklapp channels. 
\end{abstract}
\pacs{}
\vskip1pc]
\narrowtext
\section{Introduction}

The Landau theory is widely used to describe Fermi liquids even when
the interactions are strong, but it %and such a perturbative theory 
cannot be justified `a priori'. The cuprate high-T$_{\rm c}$ superconductors
show clear deviations from Landau theory in the normal state
and it has long been argued that the key to understanding these
materials lies in the breakdown of Landau theory\cite{pwa}. One possible cause
is a symmetry-breaking instability such as magnetic order. But in
experiments on underdoped cuprates\cite{timusk}, the marked deviations from Landau
theory, such as the onset of the spin gap and gaps in the ARPES
spectra near the saddle points of the Fermi surface (FS), appear without an
obvious symmetry-breaking. This raises the question whether a
breakdown of Landau theory without symmetry-breaking is possible.
Actually one example is well known and understood, the insulating spin
liquid states of even-leg ladder systems at half-filling, which have
only short range magnetic order and unbroken translational symmetry\cite{dagotto,balents,fisher,rice}.
The key to this behavior are elastic Umklapp scattering processes across the
FS which open up a charge gap at half-filling in addition to
the spin gap caused by the pairing instability. In this paper the role of
these processes in a two-dimensional system will be carefully
examined.

Renormalization group (RG) methods allow an analytical treatment and,
although the one-loop approximation is in principle applicable only at
weak coupling, we can hope to learn about possible instabilities at the
strong to intermediate couplings that apply in the cuprates. Such
methods have long been successfully applied to one-dimensional
models. The first attempts\cite{dzialoshinski,schulz,lederer}  to extend this analysis to two dimensions
were made shortly after the discovery of high temperature
superconductivity. They focussed on the dominant role of scattering
processes involving the Fermi surface regions in the vicinity of van
Hove singularities. 

Limiting the two-dimensional FS to just two patches reduces
the problem to the flow of a small number of coupling constants which
can be handled analytically. For repulsive interactions there are two
possible fixed points involving flows either to weak coupling or to
strong coupling. The possible relevance of the latter to the cuprates
was emphasized by three of the present authors\cite{furukawa}. They showed that under
certain conditions the local charge compressibility flowed towards
zero, indicating that here too Umklapp scattering opened up a local
charge gap.

A proper treatment requires that the flow of interactions involving the
whole two-dimensional Fermi surface be included. Already several RG 
investigations using a discretization of the Fermi surface into $N$
patches with $N \lesssim 32$ have been made. Zanchi and Schulz\cite{zanchi} studied 
the RG-flows of a 32-patch weak coupling Hubbard model with only
nearest-neighbor (n.n.) hopping in the kinetic energy term. They found 
a crossover between an antiferromagnetic (AF) ordered groundstate to
a $d_{x^2-y^2}$-paired superconducting (SC) groundstate as the
electron density was lowered away from half-filling. Recent more
extensive results by Halboth and Metzner\cite{halboth} have largely confirmed the
Zanchi-Schulz results, extending them to the case where there is a small
next-nearest-neighbor (n.n.n.) hopping as well and investigating possible 
incommensurate AF-orderings. Although in both these
investigations Umklapp scattering was included, the possibility of a
fixed point behavior which would be similar to that of the two-leg
ladder was not explicitly considered.

In this paper we will use a one-loop RG method with a discretization of
the FS into $N$-patches $(N=32-96)$ to examine the flow of the
coupling constants and susceptibilities under various starting
conditions. Throughout we take a substantial value for the n.n.n.
hopping amplitude, $t'$. On the one hand this is a realistic value for
the cuprates. Secondly it moves the critical density, where the
saddle points are at the FS, away from half-filling so that the saddle point effects
are not mixed with nesting effects on the zone diagonals, as
occurs when one sets $t'\equiv 0$. When $t'$ is substantial, one can
distinguish three density regions. The simplest is the strongly
doped region where the saddle points lie above the Fermi energy
and Umklapp scattering is unimportant. Here the leading instability is
to $d$-wave SC --- a form of Kohn-Luttinger instability, in agreement
with previous studies. We call this the {\em $d$-wave dominated regime}. A second relatively straightforward density
regime is the weak doping regime close to half-filling, where the
approximate nesting of FS segments near the zone diagonals dominates
and a AF instability is favored --- again in agreement with previous
studies. We call this region the {\em approximate nesting regime}. 

The intermediate regime is most interesting and will be the focus of
this work. In this case the saddle points lie slightly below the Fermi
energy and Umklapp processes involving these FS regions are highly
relevant. We call this density region the {\em saddle point regime}. As in the case of the two-leg ladder these Umklapp processes 
act to reinforce $d$-wave pairing so that this channel competes
strongly with the AF instability. If one looks only at these two
instabilities in the one-loop RG, it is not possible to decide which
dominates. In the case of the two-leg ladder the uniform (Pauli) spin
susceptibility flows to zero indicating that pairing instability
prevails --- a result confirmed when bosonization methods are used to
examine the strong coupling state below the critical scale in the
one-loop scheme. Similarly we find in the present case that an examination of the
uniform susceptibility favors an assignment of the strong coupling
fixed point to the class of the two-leg ladder. Further the local
charge compressibility defined for these FS segments appears to scale
also to zero. Below we present a detailed examination of this
 saddle point regime. We argue that rather than the simple crossover
between $d$-wave SC and AF order as the density varies, found by
previous authors, an interpretation in terms of the formation of an
insulating spin liquid (ISL) which truncates the FS segments near the saddle
points is justified. This ISL can be viewed as a form of $d$-wave RVB
(Resonance Valence Bond) state as in the case of the two-leg ladder.
Such a state represents a clear violation of the Landau
theory, which does not rely upon a translational symmetry breaking
mechanism.

Clearly an instability that partially truncates the Fermi surface with a
charge gap can be seen 
as a forerunner of the Mott insulating state which occurs for intermediate to
strong interactions. Since our motivation is to understand better the phase
diagram of the high-$T_c$ cuprates, we are most interested in such
instabilities. However we are aware that there are other instabilities which appear in a weak
coupling theory driven by the diverging density of states (DOS) at the van Hove
points. These are the Stoner instability to ferromagnetism and, as remarked by 
Halboth and Metzner\cite{halboth}, the
Labb\'{e}-Friedel or Pomeranchuck instability from square to rectangular symmetry. These split the 
saddle points when the Fermi energy lies near the van Hove singularity. This
is a drawback of using a weak coupling approach to describe an intermediate to 
strong coupling problem. We will simply ignore these DOS-related instabilities
in the forward scattering channel
and concentrate on those which we believe are more relevant as weak coupling
signatures of the intermediate to strong coupling problem.

Finally, a defect of our one--loop approximation 
is that it does not lead to a description
of the strong coupling phase of the system
which it predicts. 
There are many approaches in the literature which attempt to
construct a theory of such a state that we can loosely call a lightly doped
$d$-wave RVB state\cite{pwa,fisher,lee,zhang,vojta}. Our aim here is rather different and seeks to complement
these strong coupling theories by examining the approach from the strongly
overdoped regime which behaves as a conventional Landau-Fermi liquid with an
instability towards weak-coupling $d$-wave superconductivity. The question we
seek to address is the form of the instability in a Landau-Fermi liquid which
leads to this doped RVB state. How does it differ from a simple $d$-wave
superconducting instability and how does the proximity to the Mott insulating
state at half-filling manifest itself ?

\section{The model and its Fermi surface}
The kinetic energy of the $t-t'$ Hubbard model is given by the tight-binding 
dispersion 
\begin{equation} \epsilon (\vec{k}) = -2t (\cos k_x + 
\cos k_y ) + 4 t' \cos k_x  \cos k_y  -\mu \, \label{eofk} 
\end{equation} with nearest neighbor (n.n.) hopping $t$, next nearest 
neighbor (n.n.n.) hopping $t'$ and chemical potential $\mu$. Typically we
choose $t'=0.3 t$ which yields a 
small convex curvature of the FS around $(\pi,\pi)$ 
at higher fillings.  Another essential
curve is the Umklapp surface (US) which connects the van Hove points with
straight lines. If the  FS crosses this line, two particles at the FS can be
scattered from one side of
the US 
to the opposite one in an elastic process.  
As we will see, these additional scattering channels then enhance 
the scale of the transition to a strong coupling regime. 

The initial interaction is taken to be a simple on-site repulsion
\begin{equation}\label{HubUHam}
H_{U} = U \sum_{\vec{x}} n_{\vec{x}, \uparrow}n_{\vec{x}, \downarrow},
\end{equation}
which is constant in $k$-space. The effective interaction
will develop a pronounced $k$-space structure in the RG flow.

In recent years there have been several RG approaches to the 2D Hubbard model. Schulz \cite{schulz} and Lederer et al.\cite{lederer}  studied
the RG flow of the processes connecting the saddle points emphasizing the
divergence of both AF and $d$-wave pairing correlations. 
Dzyaloshinkii\cite{dzialoshinski} discussed  the weak coupling non-Fermi
liquid fixed point of
such a model. Similar studies have been given by Alvarez et al.\cite{alvarez}
and Gonzalez et al.\cite{gonzales}. Later on, in a related formalism based on parquet equations, Zheleznyak et al.\cite{zheleznyak} examined the interplay 
between critical scales and effects of the FS curvature for a quasi 2D model
restricted to approximately flat FS faces
% of a dispersion without n.n.n. hopping ($t'=0$) 
close to
half-filling. Another study of nesting effects between flat FS segments was
given by Vistulo de Abreu and Doucot\cite{vistulo}. 
Zanchi and Schulz\cite{zanchi} presented the first fully two--dimensional
treatment, based on Polchinski's RG equation. They studied the
2D Hubbard model with $t'=0$ and found two different regimes with
dominant AF in the one and $d$-wave pairing correlations in the other. 
A more detailed analysis of the 
leading instabilities was given by Halboth and Metzner\cite{halboth} 
using RG equations for Wick ordered functions. 
In this paper we study the RG flow for the one--particle irreducible
(1PI) vertex functions for the model given by (\ref{eofk}) and  (\ref{HubUHam}), to investigate the
possibility of a novel strong coupling phase
 which is a precursor of the Mott insulating state, as suggested by the
two-patch study by Furukawa et al.\cite{furukawa}. A self-contained and
detailed discussion of the RG
technique we use here is given elsewhere\cite{hsag}.

\section{The two-patch model revisited} 
\begin{center}
\begin{figure} %[h]
\includegraphics[width=.48\textwidth]{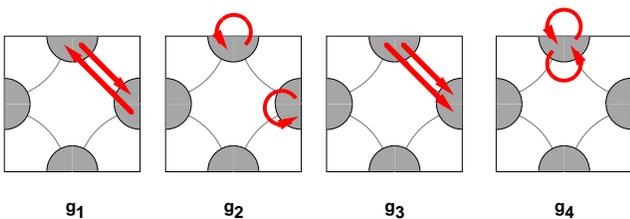} \\[0.25mm]
\caption{The relevant scattering processes in the two-patch
model. The gray semi-circles denote the phase space patches around the saddle
points. The interactions are assumed to be spin-independent and constant over
the patches. In this notation the spin of the initial and final particle
connected by an arrow has to be the same.}
\label{2patchvertices}
\end{figure}
\end{center}

\noindent
We start with a brief discussion of the dominant mechanisms
for the case %band filling close to the van Hove filling, 
where the FS is at the saddle points. These are most
transparent in the two-patch model\cite{furukawa,schulz,lederer},
where only small phase space
patches around the saddle points at $(\pi , 0)$ and $(0, \pi)$ are kept.
Neglecting possible frequency dependence we can
approximately describe the scattering processes within and between the two patches 
by four coupling constants, $g_1 \dots g_4$, 
depicted in Fig.\ \ref{2patchvertices}.
% It is important to notice that directly at the saddle points the $g_3$- and
%$g_4$-processes correspond to both Umklapp- and Cooper processes 
 The main terms which drive  the one-loop RG
flow of these vertices are:  a) the particle-particle loop $d_0$ with zero 
total incoming momentum, which diverges like $\log^2(\Lambda_0 / 
\Lambda)$ with decreasing energy scale 
$\Lambda \le \Lambda_0$ due to the van Hove 
singularity in the density of states and b) the particle-hole loop with momentum transfer $(\pi , 
\pi)$ denoted by $d_1$ which, in presence of a small but nonzero $t'$, 
diverges like $\log (\Lambda_{0} /\Lambda)$ with a large prefactor\cite{scale-fn}. 
Keeping only these two contributions, and denoting 
$y=\log(\Lambda_0 / \Lambda)$, so that decreasing $\Lambda$ means
increasing $y$, 
we obtain the RG flow equations
\begin{eqnarray} \dot{g}_1& = & 2 \dot{d_1} g_1 \, (g_2 -g_1) \, , 
\label{g1dot} \\
\dot{g}_2 &=& \dot{d_1} \, (g_2^2+g_3^2 )\, ,   \label{g2dot} \\
\dot{g}_3 &=& - 2\dot{d_0} \, g_3 g_4 + 2 \dot{d_1} \, g_3 \, (2g_2 -g_1)\, ,  \label{g3dot}  \\
\dot{g}_4 &=& -\dot{d_0} \, ( g_3^2 + g_4^2) \, . \label{g4dot} \end{eqnarray}
where $\dot g_i = \partial g_i/\partial y$, and $\dot d_0, \dot d_1 \ge 0$. 
It is useful to briefly review the analysis of the two-patch model
by Furukawa et al.\cite{furukawa}.
The second term on the right 
hand side of (\ref{g3dot})  enhances the basin of attraction of the strong 
coupling fixed point. Starting from the onsite repulsion
$g_1=g_2=g_3=g_4=U$ given by (\ref{HubUHam}), 
the coupling constants diverge at a scale $\Lambda_c$:
$g_3 \to  + \infty$, $g_4\to -\infty$ and $g_2\to + \infty$;
$g_1$ diverges more slowly. 
Initially there is a competition between the two terms on the
r.h.s.\ of (\ref{g3dot}), but the r.h.s.\ of (\ref{g4dot}) is always 
negative and thus decreases $g_4$. Eventually, $g_4$ becomes negative;
then both terms in (\ref{g3dot}) have the same sign, which accelerates 
the flow to strong coupling. 

For incoming and outgoing particles directly at
the saddle points $g_3$-processes correspond to both Cooper and
Umklapp processes. However, away from the saddle points we can distinguish
between Cooper
processes with approximately zero total incoming momentum driven through the
particle-particle channel and Umklapp
processes with momentum transfer $\approx (\pi, \pi)$ driven by the
corresponding particle-hole channel.   
>From this point of view, Eq.\ (\ref{g3dot}) states that for incoming and
outgoing wavevectors near the saddle points
the Umklapp
and the $d_{x^2-y^2}$-wave Cooper channel are coupled and mutually
reinforce each other through the $g_3$ and
$g_4$-processes which belong to both channels, thereby increasing the critical
scale $\Lambda_c$. In fact the divergence of the Umklapp
scatterings processes implies a divergence of the $d$-wave couplings and vice versa.

An analysis of the susceptibilities shows a competition between divergences
in the $d_{x^2-y^2}$-pairing and AF channel controlled by the flow of the
combinations $(g_3-g_4)$ and $(g_2+g_3)$ respectively. In this case of
competing singularities it is not clearcut which of them dominates. Furukawa et
al. proposed to resolve the issue by examining the uniform spin susceptibility 
and the charge compressibility. For a not-too-weak value of $U/t$ and $t'/t$
they found both were driven to zero by the pairing and Umklapp processes
respectively. On this basis they assigned the fixed point to be in the same
class as that of the repulsive two-leg ladder at half-filling. In that system,
the one-loop RG also exhibits 
competing and equally strong divergences in the $d$-wave pairing and 
AF channels, but the groundstate is known to be an insulating spin liquid
(ISL) from a bosonization treatment of the strong coupling
regime\cite{balents,fisher}. This form of groundstate is already signaled in the RG
calculation by the suppression towards zero in the uniform spin susceptibility 
and the charge compressibility. The ISL in the two-leg ladder at half-filling is 
a form of RVB (resonance valence bond) state with an approximate $d$-wave
pairing symmetry, but without any explicit translational or gauge symmetry breaking.

\section{The technique}
\begin{center}
\begin{figure}
\includegraphics[width=.2\textwidth]{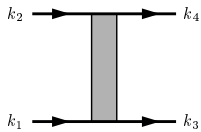}
\caption{The vertex corresponding to $V_\Lambda (\vec{k}_1,\vec{k}_2,\vec{k}_3)$}
\label{phifig1}
\end{figure}
\end{center}

\noindent
For the $N$-patch analysis we use a Wilson RG flow for one-particle irreducible (1PI) 
vertex functions. 
The full RG flow associates to every energy scale $\Lambda$ below the bandwidth
$\Lambda_0$ an effective interaction for the particles with energies 
$\epsilon(\vec k)$ below $\Lambda$, in a way that the generating functional 
for the Green functions remains independent of the scale $\Lambda$. 
Because of this exact invariance, the effective interaction is no longer just
quartic but an infinite power series in the fields. The full RG
can be expressed as an infinite hierarchy of differential equations for the 
1PI $m$--point vertex functions. 
Here we study a truncation of this infinite system in which only the 2-- and
4--point functions are kept. For a derivation of the full flow equations and a
discussion of this truncation see \cite{hsag}. Here we just state the results, which are rather simple.
Because our model is two--dimensional, continuous symmetries cannot be broken
by long--range order at any positive temperature. Therefore the effective action 
must be gauge--invariant and invariant under spin rotations, hence 
%As a consequence of these invariances, 
%In presence of spin and charge conservation, 
the four-point function is determined 
by the function 
$V_\Lambda (\omega_1,\vec{k}_1,\omega_2,\vec{k}_2,\omega_3,\vec{k}_3)$
%$V_\Lambda (\vec{k}_1,\vec{k}_2,\vec{k}_3)$ 
which describes the scattering of
two incoming 
particles $(\omega_1,\vec{k}_1,\sigma_1)$ and  $(\omega_2,\vec{k}_2,\sigma_2)$ into two outgoing
particles $(\omega_3,\vec{k}_3,\sigma_3)$ and  $(\omega_4,\vec{k}_4,\sigma_4)$ where $\sigma_1=\sigma_3$ and 
$\sigma_2=\sigma_4$, $\omega_4= \omega_1+\omega_2-\omega_3$, and $\vec{k}_4$ 
is given by momentum conservation as
$\vec{k}_4=\vec{k}_1+\vec{k}_2-\vec{k}_3$ modulo reciprocal lattice vectors. 
Because the spin of particle $1$ (first incoming) is the
same as that of particle $3$ (first outgoing), 
and similarly for $2$ and $4$, we may draw the vertex corresponding to $V_\Lambda$
as in Fig.\ \ref{phifig1}, where the solid fermion lines going through
at the top and the bottom of the vertex indicate that 
spin is conserved along these lines. 

The contributions to the right hand side of $\dot V_\Lambda = {\partial \over
\partial \Lambda } V_\Lambda$  
can then be represented graphically as in Fig.\ \ref{phifig2}.
In these graphs, 
% coupled RG of the four-point vertex and the self-energy 
one of the internal lines represents a full electron propagator 
\begin{equation}
G_{\Lambda} (\vec{k}, i \omega_n) = 
\frac{\chi_\Lambda (\vec k)}{i\omega_n - \epsilon (\vec k) 
- \chi_\Lambda (\vec k) \Sigma_\Lambda (\vec{k}, i \omega_n)}
\end{equation}
where $\chi_\Lambda( \vec{k})= 1- \left\{ \exp [ (|\epsilon|-\Lambda)/(0.05 \Lambda) ] +1 \right\}^{-1} $ cuts off energies below $\Lambda$. 
The other line stands for a single-scale
propagator %$S_{\Lambda} (\vec{k}, i \omega_n)$, which is defined as
\cite{hsag}
\begin{equation}
S_{\Lambda} (\vec{k}, i \omega_n)
=
\frac{\dot \chi_\Lambda (\vec k)\;(i\omega_n - \epsilon (\vec k)) }%
{(i\omega_n - \epsilon (\vec k) - \chi_\Lambda  (\vec k)
\Sigma_\Lambda (\vec{k}, i \omega_n))^2}
\end{equation}
Because it contains the derivative of the cutoff function with respect to $\Lambda$, 
$S_\Lambda$ is nonzero only at energies close to $\Lambda$.  
Since there are two possibilities for assigning
$G_{\Lambda}$ and $S_{\Lambda}$ to the internal lines, 
each graph stands for two contributions. 
Apart from that the usual diagrammatic rules hold: 
the graph with the fermion loop 
gets a factor $2$ from the spin trace and a minus sign. 

The contributions to the self-energy have the 
graphical representation shown in Fig.\ \ref{selfgraphs}.
Here the internal line stands for a single-scale propagator 
$S_{\Lambda}$. 

In the main part of this paper we will
neglect selfenergy corrections to the propagator.
%For the flow without self-energy corrections the full propagator is
Then $G(i\omega_n, \vec{k}) = 
\chi_\Lambda (\vec k) (i \omega_n - \epsilon( \vec{k}))^{-1}$
and the single scale propagator is simply $S_\Lambda = \partial G_\Lambda / \partial\Lambda$.
In Appendix A we show some 
results for a flow with the real part of the selfenergy on the FS taken into
account. A more complete study including selfenergy effects, in particular
the effects of the wave function renormalization, is underway. 

\begin{center}
\begin{figure} %[h]
\includegraphics[width=.36\textwidth]{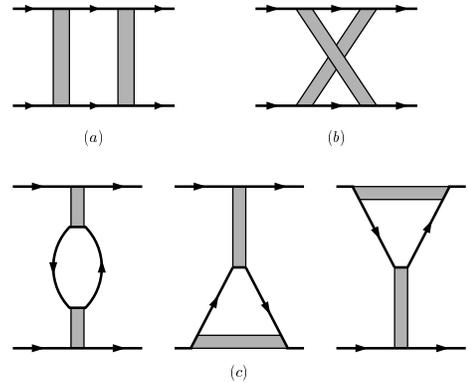}
\caption{The contributions to the right--hand side of the RGDE. 
$(a)$ the particle--particle term
$(b)$ the crossed particle--hole term
$(c)$ the direct particle--hole terms; the first of these three graphs gets 
a factor $-2$ because of the fermion loop.}
\label{phifig2}
\end{figure}
\begin{figure} %[h]
\begin{center}
\includegraphics[width=.34\textwidth]{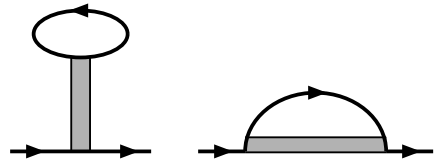}
\end{center}
\caption{The contributions to the selfenergy}
\label{selfgraphs}
\end{figure}
\end{center}

We want to emphasize that our RG method does not rely on any form of scale
invariance or scaling ansatzes. 
The RG we set up in Ref.\ \onlinecite{hsag} provides an exact rewriting of 
the generating functional in terms of the effective action. 
The approximations 
we make in the present paper are 
(i) we discard the 1PI $m$--point functions with $m \ge 6$, 
(ii) we project the four--point function to the Fermi surface and frequency zero (see the next section),
and (iii) we neglect the selfenergy corrections. 
We discuss the justification of (i) and (ii) in detail in \cite{hsag}.
The justification for (ii) is a standard RG argument, and the full momentum 
dependence can in principle be reconstructed by calculating susceptibilities and related 
quantities. The justification of (i) is less trivial but possible for 
curved Fermi surfaces and in a specific scale range \cite{hsag}.
(iii) is an approximation on which we shall improve in a further paper. 

In systems with a Cooper instability, the flow always tends towards 
strong coupling at a sufficiently low scale. This happens even in 
repulsive systems because of the Kohn--Luttinger effect. 
However, in repulsive, and initially weakly coupled systems,
the flow stays in the weak coupling regime down to a very low scale
which may never be reached because the temperature, which acts as 
a natural infrared cutoff, stops the flow before that
(in the usual Kohn--Luttinger effect, 
this scale is at most of order $e^{-\mathrm{const.}/U^2}$; see Refs.\ \onlinecite{KL,FKST}).  
In this case the system stays weakly coupled above a certain temperature,
and components of the Fermi surface limits of the four-point function 
can be identified with the Landau
interaction function $f(\vec{k},\vec{k}')$\cite{metzner,dupuis}.

When the four--point function flows to strong coupling, 
the critical scale $\Lambda_c$ where the coupling constants diverge
gives an estimate for the scale where the quasiparticles will be
strongly modified or entirely destroyed (e.g.\ for the superconducting 
transition, a gap opens up). 

The general picture we find in this model is that the flow always 
tends to strong coupling, but that for fillings where Umklapp scattering 
is favored by the geometry of the Fermi surface, the critical
scale is strongly enhanced. 

We note that in our RG method the temperature $T$ is retained 
as a physical parameter and that the RG procedure of decreasing 
the scale $\Lambda$ is a priori not related to changing the temperature.
The four--point vertex at scale $\Lambda$
is the effective interaction for the modes with energy below 
$\Lambda$, and at the same time it is the four--point function with 
infrared cutoff $\Lambda$. For $\Lambda \ll T$, one should use
the second interpretation. Note that because we do not impose a
scale-dependent cut-off on the frequencies the flow does not stop 
exactly at the point when $\Lambda$ decreases below the smallest
fermionic Matsubara frequency, $\omega_1=\pi T$. 

\section{Numerical implementation}
Next we describe the practical implementation of this RG scheme  for the 2D Hubbard model.  First we define 
a phase space discretization following Zanchi and Schulz \cite{zanchi}.  The idea is to 
discretize the Brillouin zone (BZ) into $N$ segments centered around $N$ lines.  
Each line with index $k\in \{ 1\dots N\}$ starts from the 
origin($\Gamma$-point) in a certain angular direction and 
from the $Y$ point $(\pm \pi ,\pm \pi)$ so that the lines meet at the Umklapp 
surface.  
All phase space space integrations with measure $d^2k/(2 \pi)^2$ are performed
approximately as sums over the lines and integrations over the radial 
direction.  These imply Jacobians for polar coordinates with respect 
to the $\Gamma$- or the $Y$-point, respectively.
\begin{center} 
\begin{figure} 
\includegraphics[width=.34\textwidth]{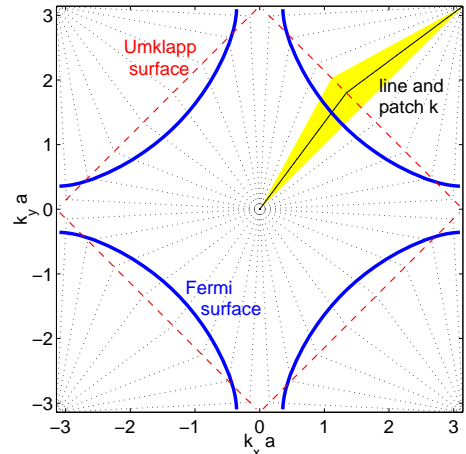}
\caption{The Brillouin zone, Fermi and Umklapp surface and the lines in the
patch centers for $N= 32$.}
\label{geom}
\end{figure}
\end{center} 
The interaction vertex $V_{\Lambda} (\vec{k}_1,\vec{k}_2,\vec{k}_3)$ depends on two incoming
wavevectors $\vec{k}_1$ and $\vec{k}_2$ and one outgoing
wavevector $\vec{k}_3$ lying in segments $k_1$, $k_2$ and $k_3$ respectively (here we have already projected the frequencies to zero). 
The fourth wave vector $\vec{k}_{4}$ is fixed by momentum conservation. In a next approximation we select a large but finite number of coupling
constants representative for certain regions in the space spanned by
$\vec{k}_1$, $\vec{k}_2$ and $\vec{k}_3$ \cite{expan-fn}. 
We choose to take these wave vectors as the crossing points of the lines $k_1$, $k_2$ or 
$k_3$ with the Fermi surface (FS), i.e. $\vec{k}_F (k_1)$, $\vec{k}_F(k_2)$
and $\vec{k}_F (k_3)$, which lie at the centers of the corresponding FS patches. By Taylor expansion 
and power counting arguments\cite{hsag}, the 
leading part of the flow is given by the coupling functions on the Fermi 
surface and at zero frequency.  Thus we approximate\cite{32x3-fn} the function
$V_{\Lambda}(\vec{k}_1,\vec{k}_2,\vec{k}_3)$ by $
V_{\Lambda}(\vec{k}_F(k_1),\vec{k}_F (k_2),\vec{k}_F (k_3) )= V_{\Lambda} (k_1,k_2,k_3)$ for all wave vectors $\vec{k}_{i}$ in the same patch 
$k_{i}$, where $i=1,2,3$.

\section{Parameters}
The initial condition for the flow of the couplings is
given by  Hubbard interactions $V_{\Lambda_0} (k_1,k_2,k_3)=U$. For 
most results discussed here we take $U=3t$. 
We choose this rather strong initial interaction because 
we are interested in the breakdown of the 
Landau-Fermi liquid due to interaction effects and do not aim at a
classification of possible weak coupling instabilities. For the most results shown here $t'=0.3t$, which is in the range of the values
reported for the cuprates.

 We will vary mainly the temperature $T$ and the particle density near and below half
filling via the chemical
potential $\mu$. We considered values between $\mu=-0.7t$ and $\mu=-1.35t$
which corresponds to fillings between $\approx 99\%$ and $\approx 62\%$ of
half-filling.  The van Hove filling, where the FS exactly touches the saddle
points, is given by $\mu =-4t'=-1.2t$. 

For a given $\mu$  the dependence of the average particle number on $T$ is
weak and irrelevant for the results. Both parameters $\mu$ and $T$ change the effective phase space for the various 
scattering processes. 
In particular, increased temperature provides a larger phase space
for particle-hole processes with momentum transfer $(\pi,\pi)$, which 
play an important role. This is similar to the quasi-1D organic conductors where
above a certain temperature the band curvature due to interchain hopping
becomes irrelevant and  1D nesting effects determine the behavior of the system\cite{bourbonnais}. 

Typically we start at energy scale $\Lambda_0 = 4t$ and integrate the flow down
to the critical scale $\Lambda_c$ which we define as the scale where the
first coupling reaches a high value like $50t$. 

\section{Couplings and susceptibilities}
We will discuss the results of our numerical RG scheme by analyzing  
the interaction on the FS and susceptibilities. In the
flow to strong coupling we will identify the most relevant, i.e. divergent couplings. In the
absence of scale invariance we cannot expect to obtain simple expressions for
the form of their divergence, therefore we use their numerical values as 
function of the scale to make a qualitative comparison. 

Along with the interactions we calculate the $d$-wave pairing
susceptibility $\chi_{dw}$ for zero pair momentum and the spin
susceptibility $\chi_s(\vec{q})$ around $\vec{q}=(\pi,\pi)$. The method we use
is described elsewhere\cite{hsag}. Typically both $\chi_{dw}$ and
$\chi_s(\vec{q})$ grow strongly as the interactions flow to strong coupling.
In general the ratio of these susceptibilities is a complicated,
%In general the competition between these susceptibilities is complex and even
non-monotonic, function of the scale. 
Therefore we do not attempt to draw sharp boundaries between different cases. 

As discussed earlier in Sec.I it is often useful to analyze also the response the
coupling to uniform external charge and spin fields
given by
\[ H_{c/s} = \int \frac{d \vec{k}}{(2 \pi) ^2}  h_{c/s} (\vec{k})  \,   (c_{\vec{k},
\uparrow}^{\dagger}c_{\vec{k}, \uparrow} \pm   c_{\vec{k},
\downarrow}^{\dagger}c_{\vec{k}, \downarrow} ) , \]
where the subscripts $c$ and $s$ stand for charge and spin respectively.
The $\vec{k}$-independent bare couplings $h_{c/s}^0$ develop a
$\vec{k}$-dependence  due to vertex renormalizations depending on $\vec{k}$
leading to dressed vertex functions $h_{c/s} (\vec{k})$. 
We cannot directly incorporate these
renormalizations of the uniform external fields in our present RG scheme 
with IR cut-off as they
involve only excitations in a low energy region of width $T$ around the
FS. Therefore we calculate $h_{c/s} (\vec{k})$ using the 
RPA for the effective theory below the IR cut-off
$\Lambda$. This means that we use the scale-dependent interactions $V_{\Lambda}
(\vec{k}_1,\vec{k}_2,\vec{k}_3)$ in the one-loop vertex corrections for the
external couplings, which then can be summed up and solved for $h_{c/s}(\vec{k},\Lambda)$ (see Appendix A).

\section{Results: three regimes}
\begin{center} 
\begin{figure}
\includegraphics[width=.48\textwidth]{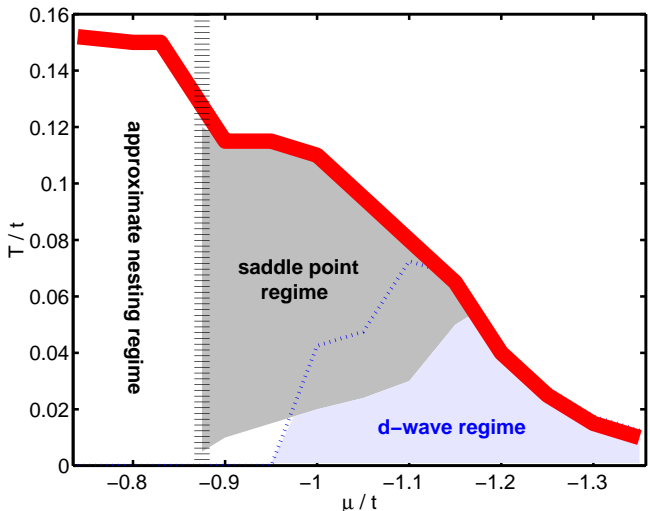}
\caption{Dependence of the flow to strong coupling on the chemical potential $\mu$ and
temperature $T$ for $t'=0.3t$ and initial interaction $U=3t$. Above the thick line we can integrate the
flow down to zero scale without reaching an instability. Below the thin broken 
line the $d$-wave pairing susceptibility $\chi_{dw}$ exceeds the AF
susceptibility $\chi_s (\pi, \pi)$ when
the largest couplings have reached the order of the band width. Above this
line, $\chi_s (\pi,\pi)$ is larger than $\chi_{dw}$. The darker gray region denotes
the {\em saddle point regime} where the charge coupling of the saddle
point regions goes to zero and the total charge compressibility is
suppressed. The lightly shaded region represents the {\em d-wave dominated
regime}. Left to the thick vertical line the instability is increasingly
dominated by couplings away from the saddle points, we refer to this region as the
{\em approximate nesting regime}.} 
\label{pdtp3}
\end{figure}
\end{center}

In the density range we examined, we always found a flow towards strong
coupling at sufficiently low temperature. The character of this divergence
of the coupling functions varies continuously  with density
and temperature. 
However we can identify three qualitatively different regimes,
which we will call the {\em $d$-wave
dominated regime}, the {\em saddle point regime} and the {\em approximate nesting
regime}, as illustrated in Fig.\ \ref{pdtp3}. 
Our analysis does not allow us to draw sharp boundaries between
the different regions. Rather, the character of the strong coupling flow
changes in a crossover--like fashion as one moves from one region 
into the other.  
%In our analysis the boundaries
%between these  appear as broad crossovers in the character of the
%strong coupling flow and can only be approximately 
%related to actual transitions
%or crossovers of the physical system. 

\begin{center} 
\begin{figure}
\includegraphics[width=.48\textwidth]{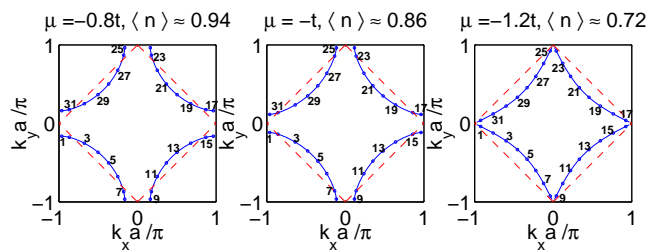} \\[1mm]
\caption{Fermi surfaces and the 32 points for the three different chemical
potentials discussed in the text. $\langle n \rangle $ denotes the average particle number per site,
 i.e. $\langle n \rangle = 1$ corresponds to half-filling. The dots on 
the FS (solid line) indicate the patch centers with patch indices given by the numbers. The
dashed line denotes the Umklapp surface (US).}
\label{fscomp}
\end{figure}
\end{center}

\begin{center} 
\begin{figure} 
\includegraphics[width=0.9\textwidth]{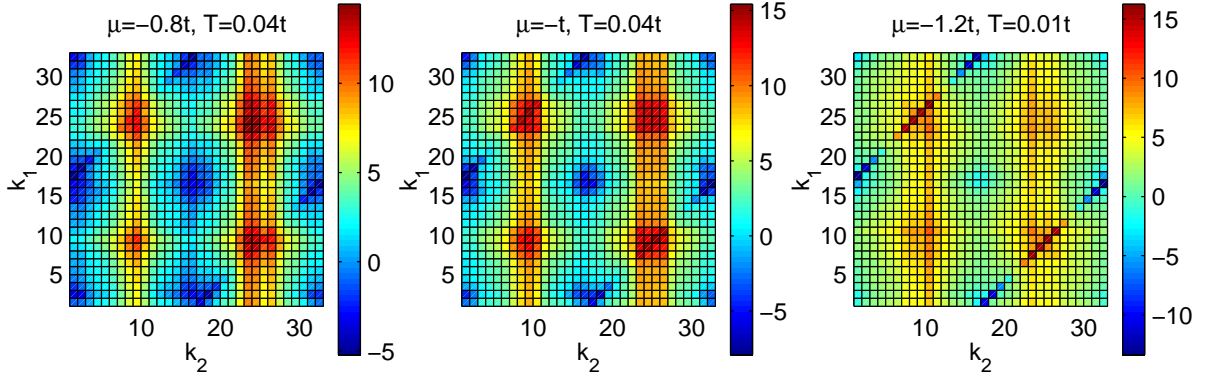}

\caption{(color) Snapshot of the couplings $V_{\Lambda} (k_1,k_2,k_3)$ with first outgoing wave vector $k_3$
fixed at point 1 (see Fig.\ \ref{fscomp})  when the largest couplings have exceeded the order of the
bandwidth for the three different choices of chemical potential and
temperature discussed in the text. The colorbars indicate the values of the couplings.}
\label{vcomp1}
\end{figure}  
\end{center}

\begin{center} 
\begin{figure} 
\includegraphics[width=0.9\textwidth]{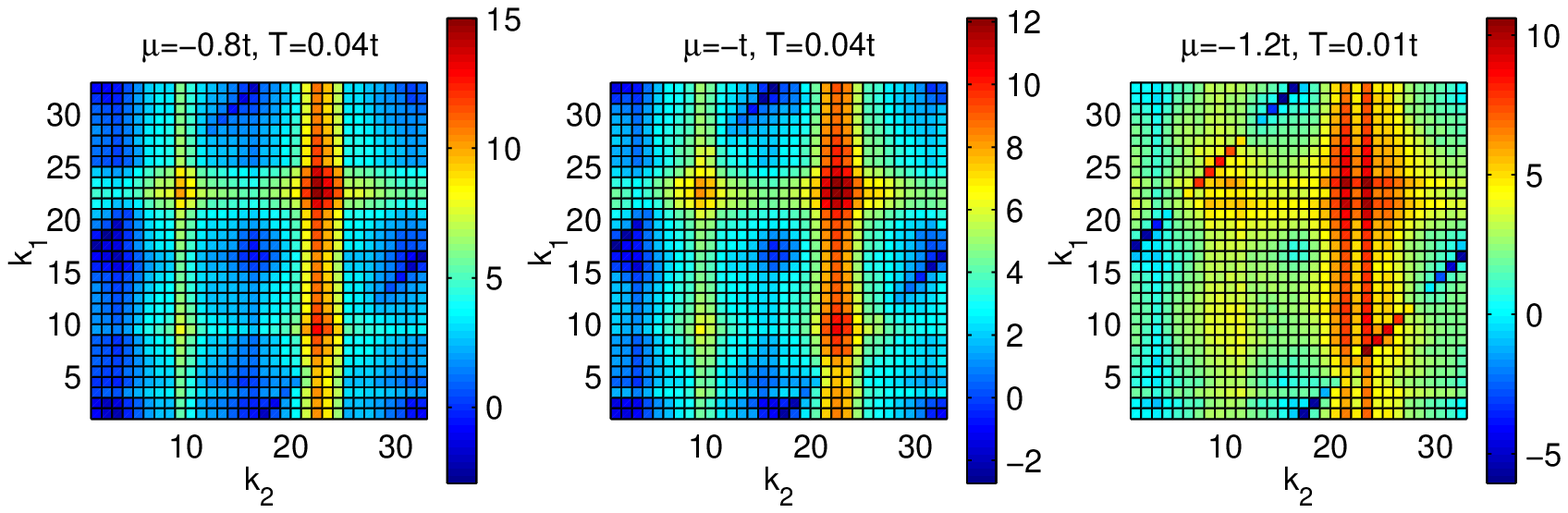}

\caption{(color) Snapshot of the couplings $V_{\Lambda} (k_1,k_2,k_3)$ with first outgoing wave vector $k_3$
fixed at point 3 (see Fig.\ \ref{fscomp}) when the largest couplings have exceeded the order of the
bandwidth for the three different choices of chemical potential and
temperature discussed in the text. The colorbars indicate the values of the
couplings. For $k_2=22$, $\vec{k}_2-\vec{k}_3 \approx 
(\pi,\pi)$ for $\vec{k}_2$, $\vec{k}_3$ close to the US.}
\label{vcomp3}
\end{figure}  
 \end{center}

\begin{center} 
\begin{figure} 
\includegraphics[width=0.9\textwidth]{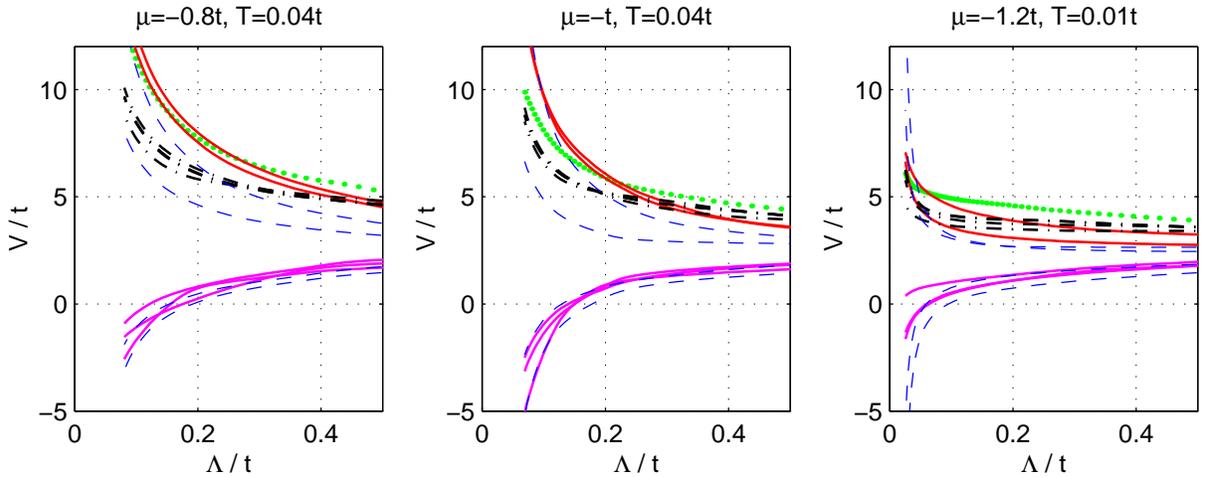}

\caption{(color) Flow of the couplings for 32-patch system: $d$-wave Cooper (blue dashed lines), $g_3$ Umklapp
(e.g. $V_{\Lambda}(24,24,1),V_{\Lambda} (23,23,2)$, solid red lines), $g_2$ forward (e.g. $V_{\Lambda}(24,1,24),V_{\Lambda}(23,2,23)$, black 
 dashed dotted lines), $g_4$
Umklapp couplings (e.g. $V_{\Lambda}(16,17,1)$, solid violet lines) and Umklapp
scatterings $V_{\Lambda}(21,21,4)$ in the BZ diagonal (green) for the three different
choices of chemical potential and temperature discussed in the text.}
\label{fvcomp}
\end{figure}  
\end{center}

\begin{center} 
\begin{figure} 
\includegraphics[width=.48\textwidth]{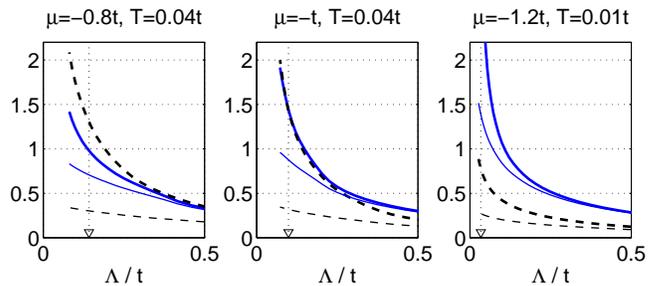}
\caption{$d$-wave (heavy solid line) and AF susceptibility(heavy dashed line)
for the three different
choices of chemical potential and temperature discussed in the text. The thin lines denote the flow
of the bare susceptibilities without vertex corrections. The mark at
the $\Lambda$-axis indicates the scale, where the largest coupling reaches $10t$.}
\label{suscomp}
\end{figure}  
\end{center}

In order to show the main
features we examine the flow for three densities typical for each regime. 
The Fermi surfaces and locations of the patch centers
that label our coupling constants
% for which the vertices are defined 
are displayed in Fig.\ \ref{fscomp}. 
In Figs.\ \ref{vcomp1} and \ref{vcomp3} we show snapshots of the 
couplings at the scale where the largest coupling have exceeded the order
of the bandwidth: 
%in the asymptotics of the flow: 
we plot the dependence of 
the coupling $V_{\Lambda} (k_1,k_2,k_3)$ with the first outgoing wave vector 
$k_3$ fixed at point 1 closest to the saddle points or at point 3 closer to  the BZ diagonal.  In 
Fig.\ \ref{fvcomp} we compare the flow of several relevant couplings 
as a function the RG scale and in Fig.\ \ref{suscomp} we plot the 
behavior of the $d$-wave pairing susceptibility $\chi_{dw}$ and the AF 
susceptibility $\chi_s( \pi,\pi)$. In the following we describe the three
regimes in detail.

\paragraph{The $d$-wave dominated regime:}
%First let us look a
At band
fillings around the van Hove filling $\mu=-1.2t$ and low temperature $T=0.01t$ 
(see right plots in Figs.\ \ref{vcomp1}, \ref{vcomp3}, and \ref{fvcomp}),
the divergence of the coupling functions only occurs at a low scale and the $d$-wave 
pair scatterings are by far the most strongly divergent couplings. 
In Fig.\ \ref{vcomp1}, they appear as red and
blue %(diverging to $\pm \infty$) 
features along the lines with patch numbers
$|k_1 - k_2| = N/2$ (on these lines, the incoming pair momentum is zero).
Other couplings like Umklapp and forward
scatterings (the red, violet and black lines in Fig.\ \ref{fvcomp}) grow, too,
but are much smaller than the Cooper couplings. This is the typical flow to
strong coupling in the lightly shaded regions in Fig.\ \ref{pdtp3}. At low
temperatures it extends also to densities slightly higher than the van Hove density.

 A closer
analysis shows that  the $d$-wave component in the  pair scattering is
generated at
intermediate scales by the particle-hole processes with momentum transfer
$(\pi,\pi)$ corresponding to the second term in Eq.\ (\ref{g3dot}). This type
of flow to strong coupling can be considered as a Kohn-Luttinger-type Cooper
instability where the repulsive scattering in the particle-hole between the saddle points first
generates a sizable initial value for the $d$-wave pair scattering and is
then gradually cut off at lower scales because the phase space for the
$(\pi,\pi)$-particle-hole processes decreases due to the shape of the FS. 

The dominance of the $d$-wave Cooper scattering is also seen in the comparison 
of the susceptibilities: the $d$-wave pairing susceptibility $\chi_{dw}$
grows much faster than the  AF susceptibility $\chi_s(\vec{q})$ (see
Fig.\ \ref{suscomp}). 

\begin{center} 
\begin{figure} 
\includegraphics[width=.48\textwidth]{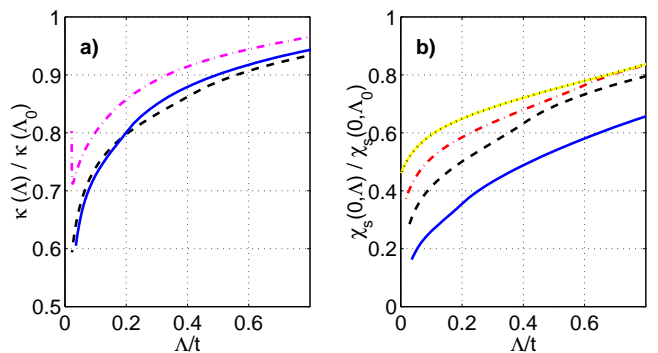}
\caption{a) Flow of the charge compressibility $\kappa$ normalized to their value at
the initial scale $\Lambda_0=4t$ for $\mu=-1.2t$ (dashed dotted line),
$\mu=-t$ (solid line) and $\mu=-0.8t$ (dashed line). b) 
Flow of the uniform spin susceptibilities normalized to their initial values
for $\mu=-t$ (solid line), $\mu=-0.8t$ (dashed line),
$\mu=-0.6t$ (dashed-dotted line), and $\mu=-0.4t$ (dotted line). For
increasing electron density, 
$\chi_s(0,\Lambda)/\chi_s(0,\Lambda_0)$ is less suppressed at low scales. For all
curves $T=0.04t$.}
\label{kachicomp}
\end{figure}  
 \end{center}

The uniform charge susceptibility $\kappa$ is somewhat
suppressed at intermediate scales but very close to the instability the
attractive Cooper scatterings in the forward scattering channel start
to dominate the vertex corrections to the charge coupling and cause a pole in
the RPA-like expression (see Eq.\ \ref{heq}) for $h_c(\vec{k})$ for $\vec{k}$
near the saddle points. This is then
the reason of a sharp upturn in $\kappa$ (see Fig.\ \ref{kachicomp}) as also observed by Halboth et
al.\cite{halboth}. At low scales the uniform spin susceptibility $\chi_s(0)$ is suppressed to 
zero by the strong attractive $g_4$ couplings favoring singlet
formation. However at higher scales, which are not related to the flow to strong coupling, the naive Stoner criterion for ferromagnetism is fulfilled due the large DOS around the van Hove filling. As discussed in the introduction, we ignore this effect. \\[2mm]

\paragraph{The saddle point regime:} 
Next we increase the temperature to $T=0.04t$ and choose
a band filling slightly above the van Hove filling such that the FS
crosses the US (chemical potential $\mu =-t$). Now the scale where the couplings reach the order of the bandwidth is
strongly enhanced. In Fig.\ \ref{vcomp1} we observe that next to the 
$d$-wave pair scatterings new features have developed. The strongly
repulsive interactions, for instance $(k_1,k_2) \approx (24,25) \to (k_3,k_4) \approx (1,17)$, correspond to $g_3$-type Umklapp scatterings which now diverge together with
the repulsive Cooper couplings. The forward scatterings of $g_2$-type also
show a strong increase towards the divergence. In addition there is a
general increase for couplings with momentum transfer $(\pi, \pi)$ 
due to the enhanced influence of the particle-hole channel with this momentum
transfer. 
On the other hand we also observe strongly attractive couplings emerging e.g.
$(k_1,k_2) \approx (16,17) \to (k_3,k_4) \approx (1,18)$. 
These processes correspond to Umklapp $g_4$ processes of pairs with both incoming
particles at the same saddle-point and outgoing particles on opposite
sides of the FS. Since these pairs have small total
momentum they couple into the Cooper channel and are driven to strong
attraction along with the attractive Cooper couplings with zero pair momentum.
This clearly demonstrates that Umklapp  and Cooper channel are strongly
coupled.
For this choice of parameters the AF susceptibility grows considerably
towards the divergence and is as large as the $d$-wave pairing
susceptibility.   

\begin{center}    
\begin{figure} 
\includegraphics[width=.48\textwidth]{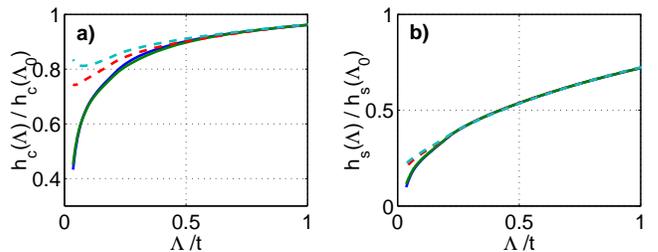}
\caption{Change of the charge $h_c(\vec{k})$ (left plot a)) and spin couplings
$h_s(\vec{k})$ (b)) normalized to their initial values of quasiparticles with wave vector
$\vec{k}$ on the FS as the electronic interactions flow to strong
coupling from a 96 point calculation at $\mu = -t$ and $T=0.04t$. The
different lines are for points close to the saddle points (solid lines), and
points closer to  the BZ diagonal (dashed lines).}
\label{usus}
\end{figure}  
 \end{center}

We call this the {\em saddle point regime} 
because the flow to strong coupling is dominated 
by the saddle point regions. Here, 
as we will show, we find the key signatures of the ISL and 
the basic mechanism of the two-patch model described in Sect.III is at work:
the diverging $g_3$-type Umklapp scattering  between the saddle point regions
drives the forward scattering of $g_2$-type to strong repulsion,
correspondingly the coupling $h_c(\vec{k})$ of external
charge fields to these FS parts and thus their contribution to the charge compressibility
$\kappa$ is increasingly suppressed as we approach the instability. This can
be seen in Figs.\ \ref{kachicomp} and \ref{usus}. In contrast to the FS near the 
saddle points
regions, which tends to incompressible i.e. insulating behavior, for
$\vec{k}$ in  the BZ diagonal the charge coupling
$h_c(\vec{k})$ is more or less unchanged. Therefore one can expect that at
wave vectors near $(\pi/2,\pi/2)$ gapless charge excitations will remain while near the
saddle points the FS will be truncated. We use this particular behavior of the $\vec{k}$-space
local charge compressibility to define the saddle point regime (darker gray regions
in Fig.\ \ref{pdtp3}): Here the charge couplings around the saddle points continue to go to zero 
if we integrate the flow far out of the perturbative range without any
indication of the upturn in $\kappa$ that we found in the
$d$-wave dominated regime. However we repeat that in this analysis the border between saddle
point and $d$-wave dominated regime is a continuous cross-over.

The uniform spin susceptibility exhibits a similar albeit somewhat more
isotropic suppression (see Figs.\ \ref{kachicomp} and \ref{usus}) when we approach 
the instability in this saddle point regime. On the one hand this is plausible
because the rapid growth of the
$d$-wave susceptibility signals strong singlet pairing tendencies. On the
other hand the AF susceptibility $\chi_s(\pi,\pi)$ seems to diverge as well, from which one
might expect long range AF order, i.e. a strong coupling state with nonzero
$\chi_s(0)$. 
\newpage
\widetext
\pagebreak \
\pagebreak
\narrowtext
Here  we argue that for the saddle point regime the more
likely candidate is a spin liquid state with strong short range AF correlation 
but nonzero spin gap. As explained, the
flow to strong coupling in this regime is caused by the coupling and mutual 
reinforcement of the $d$-wave pairing and the Umklapp processes between the broad
saddle point regions. Therefore the strong coupling state should feature
the singlet pairing of the $d$-wave channel {\em and} a strong enhancement of
$\chi_s(\vec{q})$ for $\vec{q} \approx (\pi, \pi)$. This is exactly what we observe for
$\chi_{dw}$ and $\chi_s(\vec{q})$. Moreover due to the extension of the saddle
point regions  the peak of $\chi_s(\vec{q})$ is
very broad around $(\pi,\pi)$ and does not sharpen significantly in the flow. Therefore we expect a rather short AF correlation length of 2-3
lattice spacings. This is in contrast to the $t'=0$ case very close to half
filling, where we find sharp peaks in $\chi_s (\vec{q})$ developing around $\vec{q}
= (\pi,\pi)$
and where one would
expect AF long range order at $T=0$.

\paragraph{The approximate nesting regime:}
The plots on the left  in Figs.\ \ref{vcomp1}, \ref{vcomp3}, and \ref{fvcomp} show the flow for a higher
filling ($\mu=-0.8t$).  In this case the leading interactions are Umklapp couplings between the BZ regions where
the FS intersects the US (see red features in Fig.\ \ref{vcomp3}) while the importance of the vicinity of the saddle points decreases. We call this the {\em
approximate nesting regime}. The dominating FS regions are now further away from 
the saddle points due to the higher filling. As a consequence the coupling
between Umklapp and pairing channel decreases and the $d$-wave pairing
processes become less relevant. This can be seen best from the weaker flow of the
attractive Cooper couplings in Fig.\ \ref{fvcomp}.  Now the AF susceptibility clearly
exceeds the $d$-wave pairing susceptibility. This signals increasing AF
ordering tendencies which are in accordance with sharper $(\pi,\pi)$ features
in the interactions (see Fig.\ \ref{vcomp3}), decreasing suppression of
$\chi_s(0)$ relative to its initial value (see Fig.\ \ref{kachicomp}), and a sharper peak of $\chi_s(\vec{q})$ 
around $(\pi,\pi)$. The charge susceptibility is also suppressed like in the
saddle point regime, however the FS regions with smallest charge couplings stay
fixed to the US and therefore move towards the BZ diagonal if we increase the 
filling.\\[2mm]

We emphasize that in our RG treatment the next nearest neighbor hopping $t'$ is
important for the existence of a sizable saddle point regime. For zero or very
small $t'$ the FS is closer to the US in the BZ diagonals and the $(\pi,\pi)$
scattering between the rather flat FS faces dominates even more strongly than in
our approximate nesting
regime with more FS curvature. If we now decrease the band filling, at some
point, as pointed out by Zheleznyak et al.\cite{zheleznyak} and explicitly shown for
the 2D case by Zanchi and Schulz\cite{zanchi}, these processes are cut off at low scales
and can only serve as generators of an attractive $d$-wave initial condition. With 
$t'$ very small the system crosses rather sharply from a nesting regime
into a $d$-wave dominated regime without going through
a saddle point regime in between.

Apart from the suppression of the total charge compressibility described 
above there are other potential instabilities in the forward
scattering channel. For example, as pointed out by Halboth and Metzner\cite{halboth}, there
appears to be a strong tendency towards Labb\'{e}-Friedel or Pomeranchuck FS deformations which break the square
symmetry. These are mainly rectangular deformation modes which split the 
degeneracy of the saddle-points. However we
will ignore those tendencies for the reasons discussed above. We have checked that a moderate %$d$-wave like 
deformation of the FS that breaks the square symmetry and leads to saddle point
splittings of the order of the critical scale $\sim 0.1t$ does not invalidate
the results described above. 

A difference to the two-patch analysis of Sec.III is that the saddle point
regime, where we observe the ISL signatures in our $N$-patch
calculation, 
is found at positive temperatures and densities slightly
higher than the van Hove density assumed in the two-patch analysis.
The reason for the latter is that in the $N$-patch flow the FS parts away
from the saddle points reinforce mainly the Cooper channel. Only if
the FS really crosses the US there is sufficient low energy phase
space for the Umklapp processes which then act together
with the Cooper processes leading to the unusual strong coupling flow.
For similar reasons nonzero temperature is needed for the saddle point
regime. Moderate $T$ smears out the FS and provides additional
phase space for both the particle-particle processes with small total momentum 
and the particle-hole processes with momentum transfer $(\pi,\pi)$. Especially 
due to the latter there is a certain temperature range where this
thermal phase
space gain outweighs the ordinary decrease of the one-loop contributions for
increasing $T$ and the critical scale $\Lambda_c$ is enhanced with respect to its $T=0$ value. 
 
\section{Discussion and conclusions}
We have presented an $N$-patch renormalization group analysis of the 2D
Hubbard model and found indications that the path from a Fermi liquid-like
state to the Mott insulating state may pass through a spin liquid phase with
partially truncated FS and incompressible regions around the saddle points. 
Certainly the above results have to be interpreted with care and are only qualitative as they are an attempt 
to learn about  possible strong coupling states  
from extrapolating weak--coupling flows. 
However they demonstrate that the breakdown of a Fermi liquid
through an ISL state with partially truncated FS seems to be a viable 
concept, because in the saddle point regime the qualitative features 
of the ISL, e.g. spin and 
charge gaps, are visible as tendencies in our weak coupling approach. The 
essential phenomenon which can be identified as the cause for the ISL 
in the two-patch model, namely the coupling of Umklapp and pairing 
channel,  is found to exist in a sizable temperature 
and density range also in our improved RG calculation, which includes the entire Fermi surface. 
We believe that this behaviour is robust because  
it only requires sufficiently large low energy phase space around the saddle 
points but does not rely on further details of the interaction or 
dispersion relation.  What remains to be clarified is when this 
interplay between pairing and Umklapp processes, which %intrinsically 
frustrates symmetry-breaking tendencies and thus leads to an ISL, indeed represents an energetically favorable situation for the 
system.  Another interesting and related aspect is the question of
the precise conditions for which the overlap between the channels becomes 
too small, such that at $T=0$ the system can still undergo 
a transition into a symmetry-broken state with 
presumably renormalized properties. In our calculation such symmetry-broken
states are suggested on either 
side of the saddle point regime, e.g. in the
$d$-wave dominated phase or closer to half-filling in the approximate nesting regime.

Our approach certainly
bears some appealing features when compared to the high-$T_c$ cuprates. However note that especially very close to half
filling, in the approximate nesting regime, our description will be much too simple, as
interaction effects which are not taken into account will
become large. On the other hand further away from half filling in the saddle
point regime we can hope to give a reasonable qualitative description of the 
driving forces for the breakdown of the Landau-Fermi liquid. Due to 
the mutual reinforcement between Cooper and Umklapp channel $d$-wave 
pairing correlations appear in a natural way at an enhanced scale on 
the threshold to the Mott state. If 
the insulating tendencies are strong enough they will lead to the ISL 
formation around the saddle points.

The stabilization of the ISL in the vicinity of the saddle points opens up a
novel channel to enhance Cooper pairing on the remaining open parts of the
FS. A similar mechanism was recently proposed by Geshkenbein et
al.\cite{geshkenbein} who examined a model with infinite mass preformed pairs
existing at higher temperatures also in the vicinity of the saddle points. Let 
us assume that the ISL has formed in a region (called the $A$ region) around
the saddle points at an energy scale $\Lambda_{\mathrm{ISL}}$. Then the 
dominant coupling between  the ISL and the open FS parts (called the $B$
regions) will occur through the
exchange of zero momentum hole pairs in the Cooper channel. Further it will
occur in the $d$-wave pairing channel. We denote by $V_{AB}$ the pair
scattering matrix element between the ISL in the $A$-regions and the open $B$
regions of the FS at the scale $\Lambda_{\mathrm{ISL}}$ and
by $\epsilon_A$ the energy relative to the chemical potential to add a hole
pair to the ISL. Then, at energy scales $\Lambda < \Lambda_{\mathrm{ISL}}$, an
additional attraction $V_{BB}$ is generated between pairs in
the open $B$ regions, which has a pole at
\begin{eqnarray} \Lambda_c^B &=& \Lambda_{\mathrm{ISL}}\,  \exp \left( -
\frac{\epsilon_A}{N_{AB}} \right)
\, , \label{clc} \\ N_{AB} &=& n_A \int_{\mathrm{B-\mathrm{FS}}}
\frac{dk}{(2\pi)^2} \, \frac{V_{AB}^2(\vec{k})}{v_F(\vec{k}) } \, .  \nonumber \end{eqnarray}
Here the integral is over the Fermi surface of the $B$ regions and $n_A$
denotes the number of intermediate states with two additional particles in the
$A$ regions per lattice site.  

Although we do not have a full theory of the strong coupling phase, it is
plausible to assume that in the $A$-regions a charge gap spreads out along
the US in analogy to ladder systems which when lightly doped show
simultaneously channels with and without charge gap\cite{rice,ledermann}. Then the open $B$-region of the FS will enclose an area measured from
the US, determined by the hole density and so the superfluid density will be
given by the hole density in the saddle point regime. Also there will be two
energy scales, a higher one determining the onset of the ISL in the $A$
regions, and a lower scale setting the transition temperature to
superconductivity, $T_c$. These features are in nice qualitative agreement
with the observations in the underdoped cuprates. However a full microscopic
theory of the strong coupling phase and also the crossover to the more
conventional $d$-wave superconductivity in the lower electron density $d$-wave
regime remains to be worked out. Note that in the latter regime Umklapp scattering is
irrelevant at low energy scales and the superfluid density is determined by
the electron density, not the hole density.

Finally we note that the ISL concept might provide a microscopic basis for
understanding the ARPES results on the cuprates which clearly show the
truncation of the FS around the saddle points\cite{norman} and also for phenomenological models\cite{ioffe} which have proven to be plausible 
descriptions of the transport properties of the normal state of the 
underdoped high-$T_{c}$ cuprates.

\acknowledgements
We would like to thank G.\ Blatter, J.\ Feldman, D.\ Geshkenbein, L.\ Ioffe, H.\ Kn\" orrer, W.\ Kohn, U.\ Ledermann, K.\ LeHur, W.\ Metzner, 
M.\ Troyer, E.\ Trubowitz, F.C.\ Zhang,
and M.\ Zhitomirsky for valuable discussions. 
C.H.\ acknowledges financial support by the Swiss National Fonds.

\appendix
\section{The one-loop self-energy and the Fermi surface shift}
\begin{center} 
\begin{figure}
\includegraphics[width=.48\textwidth]{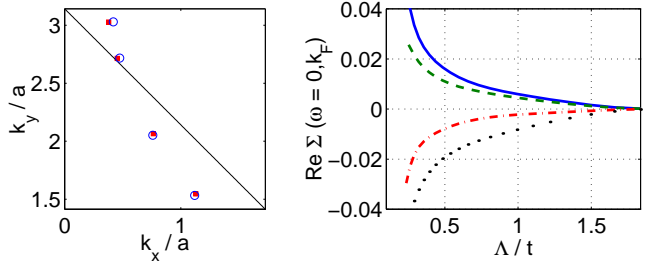}
\caption{Left: Initial (open circles) and final (squares) FS for $t'=0.3t$ and
$\langle n \rangle \approx 0.88$ per site. Right: Flow of the self-energy on the FS (solid line: point closest to 
the saddle points, dotted line: point closest to BZ diagonal). The flow was stopped when
the largest coupling reached the bandwidth $8t$.} \label{fsflow}
\end{figure}
\end{center}

\noindent
Here is a short overview of the results for the RG flow of the Fermi surface
with fixed particle number.  In order to obtain the FS flow we 
calculate in every RG step the change of the one-loop selfenergy 
given by the contributions in Fig.\ref{selfgraphs}.  Due to the approximations
made for the couplings this selfenergy is constant over a single patch and 
only yields a patch-dependent shift of the Fermi surface.   
In order to keep the particle number fixed we adjust the chemical 
potential after each step.  Quite generally we find that the FS parts 
which have the strongest repulsive scatterings with momentum transfer 
close to $(\pi ,\pi)$ develop positive self-energies and are therefore 
are shifted inwards during the RG flow.  The reason for that becomes 
clear if one considers a model interaction which is sharply peaked 
and repulsive at $\vec{Q}= (\pi , \pi )$.  For the self energy $\Sigma 
(\vec{k})$ of a particle with wave vector $\vec{k}$ on the FS one primarily 
has to examine the Hartree term (first term in Fig.\  \ref{selfgraphs}), which 
is the main contribution for the typical divergence of the couplings.  
This diagram contains a propagator with a differentiated cut-off 
function and gives a positive contribution if the state $\vec{k}+ 
\vec{Q}$ is occupied and zero contribution otherwise.  After 
subtraction of the FS average of $\Sigma(\vec{k}_F)$ (or more precisely a 
constant which keeps the particle number fixed) this yields a positive 
selfenergy $\Sigma 
(\vec{k})> 0$ for particles outside the US (because then in general the state $\vec{k}+ 
\vec{Q}$ is occupied for a FS with the densities and $t'$ values we are
interested in) and a negative shift $\Sigma 
(\vec{k}) < 0$ for states inside the US.  In our case the 
interaction only has a broad peak around $(\pi , \pi)$, therefore in 
general also FS points inside the BZ can be pushed inwards provided 
they are more affected by this repulsion than the average FS (this 
happens in the overdoped $\mu=-1.3 t$ case). The flow of the the selfenergies with 
fixed FS close to the instability is shown in Fig. \ref{fsflow} b)
 for different positions on the FS: for the FS points near the saddle points
$\Sigma_{\Lambda}  
(\vec{k})$ flows to positive values, while for $\vec{k}_F$ in the BZ diagonal
it becomes negative.  The resulting movement of the FS points if we include
$\Sigma_{\Lambda} (\vec{k})$ in the dispersion, i.e. allow the FS to move, can be seen in Fig.\ref{fsflow} a). It reveals the tendency of the FS to become flat, thus remaining in the vicinity of the 
Umklapp surface\cite{tp0-fn}. In both cases the density is kept fixed at 
$\langle n \rangle \approx 0.88$ per site. Our RG results are in qualitative
agreement with calculations using a model interaction due to AF spin fluctuations\cite{yanase}
and the FLEX approximation\cite{morita}.

\section{Calculation of the uniform susceptibilities}
\begin{center}
\begin{figure} %[h]
\includegraphics[width=.48\textwidth]{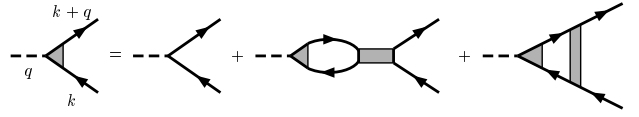}
\caption{Diagrammatic expression for the renormalization of the
couplings $h_{c/s} (\vec{k})$ to external charge or magnetic fields. For 
uniform external fields we take $\vec{q} \to 0$.}
\label{hcouplings}
\end{figure}
\end{center}

The uniform ($\vec{q} \to 0$) susceptibilities describing the response to
external charges and magnetic fields cannot be calculated successively 
by lowering
the IR cut-off as they only involve degrees of freedom very close to the FS (the 
width of this region is given by the temperature). Therefore we determine
these responses for the effective theory below the cut-off $\Lambda$ with the
interactions at this scale as the effective interactions  renormalizing the
coupling to the external fields via vertex corrections. More precisely we
calculate the effective couplings $h_i(\vec{k})$ ($i=c$ for charge and $i=s$
for spin) of quasiparticles on the FS,
occurring in the Hamiltonian as
\[\int \frac{d \vec{k}}{(2 \pi) ^2}  h_{c/s} (\vec{k}) \,   (c_{\vec{k},
\uparrow}^{\dagger}c_{\vec{k}, \uparrow} \pm   c_{\vec{k},
\downarrow}^{\dagger}c_{\vec{k}, \downarrow} ) . \]
Denoting the bare coupling as $h_{c/s}^0$ we can express the effective coupling as
\begin{equation} h_i (\vec{k})= h^0_i(\vec{k}) + \int \frac{d \vec{k'}}{(2
\pi) ^2} \, h_i(\vec{k}) \Phi(\vec{k'}) V_i(\vec{k},\vec{k'}) \, \quad i=c, \,s   \label{heq}
\end{equation} 
where $V_c(\vec{k},\vec{k'}) =
(-V_{\Lambda}(\vec{k},\vec{k'},\vec{k'})+2V_{\Lambda}(\vec{k},\vec{k'},\vec{k}))$ for the charge
and $V_s(\vec{k},\vec{k'}) = -V_{\Lambda}(\vec{k},\vec{k'},\vec{k'})$ for the spin
coupling. Diagrammatically this equation is shown in
Fig.\ \ref{hcouplings}. 
The kernel $\Phi(\vec{k})$ is the $\omega=0$, and then
$\vec{q} \to 0$ limit of the Matsubara sum of the product of two propagators,
and given by the derivative of the Fermi function:
\begin{equation}
\Phi(\vec{k}) = 
\lim_{\vec{q} \to 0} 
\textstyle{\frac{n_F(\epsilon (\vec{k}+\vec{q})) - n_F (\epsilon
(\vec{k}))}{\epsilon (\vec{k}+\vec{q})-\epsilon (\vec{k}))} }
=
\frac{d n_F}{d E} \vert_{E=\epsilon(\vec k)}.
\end{equation}
%
%\begin{eqnarray}  \lim_{\vec{q} \to 0} \frac{n_F(\epsilon (\vec{k}+\vec{q})) - n_F (\epsilon
%(\vec{k}))}{\epsilon (\vec{k}+\vec{q})-\epsilon (\vec{k}))} \; &\longrightarrow&
%\nonumber \\ -
% \frac{1}{T} \, \frac{\exp (\epsilon(\vec{k}) /T) }{ \left( \exp
%(\epsilon(\vec{k})/T) +1 \right)^2} &=& \Phi (\vec{k}) \quad \mbox{for} \; \vec{q} \to 0 \, .
%\end{eqnarray}
 The uniform susceptibilities are
then given as \begin{eqnarray} \kappa = - \int \frac{d^2k}{(2\pi)^2} \, 
h_c^0(\vec{k})\Phi (\vec{k}) h_c(\vec{k}) \, , \\ \chi_s(0) = - \int 
\frac{d^2k}{(2\pi)^2} \, h_s^0(\vec{k})\Phi (\vec{k}) h_s(\vec{k}) \, 
.  \end{eqnarray}
In absence of an instability the coupling functions for zero
momentum transfer $V_{\Lambda}(\vec{k},\vec{k'},\vec{k})-\frac{1}{2} V_{\Lambda}(\vec{k},\vec{k'},\vec{k'})$
and $-\frac{1}{2} V_{\Lambda}(\vec{k},\vec{k'},\vec{k'})$ would converge to the Landau interaction
functions $f_s(\vec{k},\vec{k'})$ and $f_a(\vec{k},\vec{k'})$,
respectively,
and the expressions for the susceptibilities obtained with the
above scheme reduce to the results from Fermi liquid theory.


\begin{thebibliography}{99}
\bibitem{scale-fn} Below an energy scale determined by $t'/t$ and $\mu$, 
these terms become small compared to $\dot d_0$, which then
dominates the flow. However if the initial coupling is not very weak,
the coupling constants diverge already above that scale.
\bibitem{expan-fn}
As compared to an expansion in basis functions, this has the advantage 
that one can easily understand the dominant processes in momentum space.
\bibitem{32x3-fn} We have also calculated the flow
including a radial dependence of $V_\Lambda$, by 
keeping three radial points in each patch (inside, on, and outside of the FS, i.e.\  $(32\times 3)^3$ coupling constants).
The main conclusions are the same as the ones without radial dependence.
\bibitem{tp0-fn} In the $t'=0$ case close to half filling the 
states in the BZ diagonal are slightly more affected by the strong 
$(\pi, \pi)$-repulsion than the states at the saddle points.  
Therefore the FS expands towards the saddle points, although the trend 
is weaker than for $t'=0.3t$.
\end{thebibliography}
\end{document}